\documentclass[twocolumn,showpacs,preprintnumbers,amsmath,amssymb]{revtex4}

\usepackage{graphicx}
\usepackage{epsfig}
\usepackage{rotating}
\usepackage{dcolumn}
\usepackage{bm}

\begin{document}

\preprint{HEP/123-qed}
\title{Photoproduction of $\pi^0$-mesons off neutrons in the nucleon resonance region
}
\author{
  M.~Dieterle$^{1}$,
  I.~Keshelashvili$^{1}$,
  J.~Ahrens$^{2}$,
  J.R.M.~Annand$^{3}$,
  H.J.~Arends$^{2}$,
  K.~Bantawa$^{4}$,
  P.A.~Bartolome$^{2}$, 
  R.~Beck$^{2,5}$,
  V.~Bekrenev$^{6}$,
  A.~Braghieri$^{7}$,
  D.~Branford$^{8}$,
  W.J.~Briscoe$^{9}$,
  J.~Brudvik$^{10}$,
  S.~Cherepnya$^{11}$,
  B.~Demissie$^{9}$,
  E.J.~Downie$^{2,3,9}$,
  P.~Drexler$^{12}$,
  L.V.~Fil'kov$^{11}$, 
  A.~Fix$^{13}$,      
  D.I.~Glazier$^{8}$,
  D.~Hamilton$^{3}$, 
  E.~Heid$^{2}$,
  D.~Hornidge$^{14}$,
  D.~Howdle$^{3}$,
  G.M.~Huber$^{15}$,
  I.~Jaegle$^{1}$,
  O.~Jahn$^{2}$,
  T.C.~Jude$^{8}$,
  A.~K{\"a}ser$^{1}$,   
  V.L.~Kashevarov$^{2,11}$,
  R.~Kondratiev$^{16}$,
  M.~Korolija$^{17}$,
  S.P.~Kruglov$^{6}$, 
  B.~Krusche$^1$,   
  A.~Kulbardis$^{6}$,  
  V.~Lisin$^{16}$,
  K.~Livingston$^{3}$,
  I.J.D.~MacGregor$^{3}$,
  Y.~Maghrbi$^{1}$,
  J.~Mancell$^{3}$, 
  D.M.~Manley$^{4}$,
  Z.~Marinides$^{9}$,
  M.~Martinez$^{2}$,
  J.C.~McGeorge$^{3}$,
  E.~McNicoll$^{3}$,
  D.~Mekterovic$^{17}$,
  V.~Metag$^{12}$,
  S.~Micanovic$^{17}$,  
  D.G.~Middleton$^{14}$,
  A.~Mushkarenkov$^{7}$,
  B.M.K.~Nefkens$^{10}$,
  A.~Nikolaev$^{2,5}$,
  R.~Novotny$^{12}$,
  M.~Oberle$^{1}$,
  M.~Ostrick$^{2}$,
  B.~Oussena$^{2,9}$, 
  P.~Pedroni$^{7}$,
  F.~Pheron$^{1}$,
  A.~Polonski$^{16}$,
  S.N.~Prakhov$^{10}$,
  J.~Robinson$^{3}$,   
  G.~Rosner$^{3}$,
  T.~Rostomyan$^{1}$,
  S.~Schumann$^{2,5}$,
  M.H.~Sikora$^{8}$,
  D.~Sober$^{18}$,
  A.~Starostin$^{10}$,
  I.~Supek$^{17}$,
  M.~Thiel$^{2,12}$,
  A.~Thomas$^{2}$,
  M.~Unverzagt$^{2,5}$,
  D.P.~Watts$^{8}$,
  D.~Werthm\"uller$^1$,
  L.~Witthauer$^1$\\
(Crystal Ball/TAPS experiment at MAMI, the A2 Collaboration)
}
\affiliation{
  $^{1}$\mbox{Department of Physics, University of Basel, Switzerland}\\
  $^{2}$\mbox{Institut f\"ur Kernphysik, University of Mainz, Germany}\\
  $^{3}$\mbox{Department of Physics and Astronomy, University of Glasgow, Glasgow, UK}\\
  $^{4}$\mbox{Kent State University, Kent, OH, USA}\\  
  $^{5}$\mbox{Helmholtz-Institut f\"ur Strahlen- und Kernphysik, University of Bonn, Germany}\\
  $^{6}$\mbox{Petersburg Nuclear Physics Institute, Gatchina, Russia}\\
  $^{7}$\mbox{INFN Sezione di Pavia, Pavia, Italy}\\
  $^{8}$\mbox{School of Physics, University of Edinburgh, Edinburgh, UK}\\
  $^{9}$\mbox{Center for Nuclear Studies, The George Washington University, Washington, DC, USA}\\
  $^{10}$\mbox{University of California at Los Angeles, Los Angeles, CA, USA}\\
  $^{11}$\mbox{Lebedev Physical Institute, Moscow, Russia}\\
  $^{12}$\mbox{II. Physikalisches Institut, University of Giessen, Germany}\\
  $^{13}$\mbox{Laboratory of Mathematical Physics, Tomsk Polytechnic University, Tomsk, Russia}\\
  $^{14}$\mbox{Mount Allison University, Sackville, New Brunswick E4L 1E6, Canada}\\
  $^{15}$\mbox{University of Regina, Regina, SK S4S 0A2 Canada}\\
  $^{16}$\mbox{Institute for Nuclear Research, Moscow, Russia}\\
  $^{17}$\mbox{Rudjer Boskovic Institute, Zagreb, Croatia}\\
  $^{18}$\mbox{The Catholic University of America, Washington, DC, USA}\\
}
\date{\today}

\begin{abstract}
Precise angular distributions have been measured for the first time for the photoproduction 
of $\pi^{0}$-mesons off neutrons bound in the deuteron. The effects from nuclear Fermi motion 
have been eliminated by a complete kinematic reconstruction of the final state. The influence 
of final-state-interaction effects has been estimated by a comparison of the reaction cross 
section for quasi-free protons bound in the deuteron to the results for free protons and then 
applied as a correction to the quasi-free neutron data. The experiment was performed at 
the tagged photon facility of the Mainz Microtron MAMI with the Crystal Ball and TAPS detector 
setup for incident photon energies between $0.45$~GeV and $1.4$~GeV. The results are compared to 
the predictions from reaction models and partial-wave analyses based on data from other isospin 
channels. The model predictions show large discrepancies among each other and the present data 
will provide much tighter constraints. This is demonstrated by the results of a new analysis in 
the framework of the Bonn-Gatchina coupled-channel analysis which included the present data.
\end{abstract}

\pacs{13.60.Le, 14.20.Gk, 14.40.Aq, 25.20.Lj
}

\maketitle

The excitation spectrum of the nucleon is generated by the strong force and thus
should reflect its basic properties. The impossibility of a perturbative treatment 
of Quantum Chromodynamics (QCD) on the relevant energy scale of a few GeV has so 
far precluded ab-initio calculations of the properties of nucleon resonances. 
However, with the recent progress in the numerical methods of lattice gauge calculations, 
such predictions have come within reach. First, unquenched lattice calculations support 
the SU(6)$\otimes$O(3) excitation structure of the nucleon, familiar from the 
constituent quark model, and second, also have a level counting consistent with the 
non-relativistic quark model \cite{Edwards_11}. Although these calculations are 
in a very early state, the result is interesting because the agreement between
the quark-model results and the experimentally established excitation spectrum of the
nucleon is not very good. For most quantum numbers, apart from the first excited 
state, there are no counterparts from experiment \cite{Arndt_06} to the plethora of 
states predicted by models. For the identified excited states electromagnetic
couplings to the nucleon ground state are a very sensitive observable for the 
test of nucleon models, because they reflect the spin-flavor correlations in 
the wave functions.  

This situation has triggered large efforts to improve the experimental data base 
for nucleon resonances exploiting photon-induced meson production reactions.
Due to the progress in accelerator and detector technology, such reactions can now be 
studied with comparable or even better precision than hadron-induced reactions, although 
the typical cross sections are smaller by roughly three orders of magnitude. 
Although many different final states are under investigation to avoid bias from the 
coupling of the excited nucleon states to specific decay channels, pion production
always was and still is a cornerstone for the extraction of nucleon resonance
properties such as masses, widths, and electromagnetic couplings
\cite{MAID,MAID_new,SAID,SAID_new,BnGa,DMT,Feuster_99,Fernandez_06,Aznauryan_09,Shresta_12}. 
In photoproduction reactions, neutral pions are of special interest because 
they do not couple directly to photons so that non-resonant background contributions 
are suppressed. 

Previous experimental efforts for $\pi^0$ production have concentrated on the measurement 
of angular distributions and polarization observables for the free proton target
\cite{Bartholomy_05,Bartalini_05,vanPee_07,Dugger_07,Elsner_09,Sparks_10,Crede_11,Thiel_12,Gottschall_14,Sikora_14}. 
The recent measurements of the double polarization observables $G$ (linearly 
polarized beam, longitudinally polarized target \cite{Thiel_12}), $E$ (circularly
polarized beam, longitudinally polarized target \cite{Gottschall_14}), and $C_x^{\star}$
(circularly polarized beam, polarization of recoil nucleon \cite{Sikora_14}) for this reaction 
had significant impact on the analysis of nucleon resonance properties. They
are important steps towards a `complete' measurement, which allows a unique, model-independent
extraction of the reaction amplitudes \cite{Chiang_97}.

However, the isospin decomposition of pion photoproduction requires also measurements
with neutron targets, which can only be done with quasi-free neutrons bound in a light
nucleus, in most cases the deuteron. The data base for such reactions is so far scarce.
For isovector mesons like pions three independent amplitudes contribute to photoproduction, 
the isoscalar part $A^{IS}$, the isospin dependent amplitude $A^{IV}$, and the isospin changing
amplitude $A^{V3}$, which are related by \cite{Walker_69}:  
\begin{eqnarray}
\label{eq:iso}
A(\gamma p\rightarrow\pi^+ n) & = &
-\sqrt{\frac{1}{3}}\;A^{V3}+\sqrt{\frac{2}{3}}(A^{IV}-A^{IS}) \\
A(\gamma p\rightarrow\pi^o p) & = &
+\sqrt{\frac{2}{3}}\;A^{V3}+\sqrt{\frac{1}{3}}(A^{IV}-A^{IS})\nonumber\\
A(\gamma n\rightarrow\pi^- p) & = &
+\sqrt{\frac{1}{3}}\;A^{V3}-\sqrt{\frac{2}{3}}(A^{IV}+A^{IS})\nonumber\\
A(\gamma n\rightarrow\pi^o n) & = &
+\sqrt{\frac{2}{3}}\;A^{V3}+\sqrt{\frac{1}{3}}(A^{IV}+A^{IS})\;.\nonumber
\end{eqnarray}
Isospin $I=3/2$ $\Delta$ resonances are only excited by $A^{V3}$, so that their electromagnetic 
couplings are identical for protons and neutrons, while $I=1/2$ $N^{\star}$ states couple 
differently to protons and neutrons.   
\begin{figure}[thb]
\centerline{\resizebox{0.50\textwidth}{!}{%
  \includegraphics{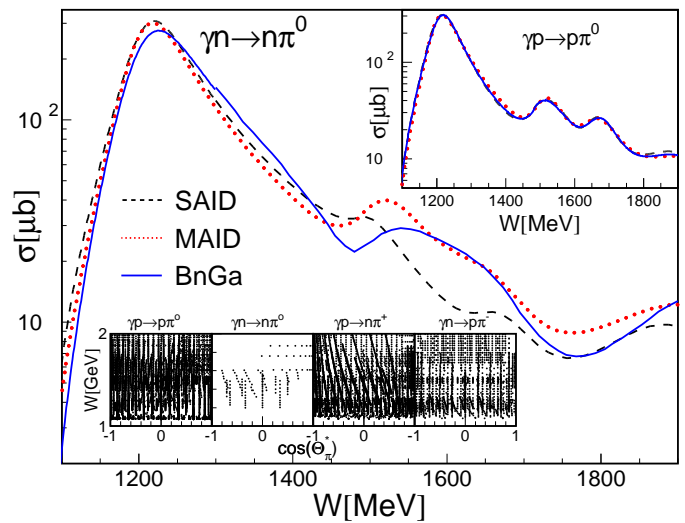}
}}
\caption{Main plot: predicted total cross sections from SAID \cite{SAID_new}, 
MAID \cite{MAID_new}, and the BnGa analysis \cite{BnGa_n} for $\gamma n\rightarrow n\pi^0$.
Insert, upper right corner: same analyses for $\gamma p\rightarrow p\pi^0$.
Insert at bottom: previously available data base \cite{Krusche_11} (each point represents 
one measurement at $W$ and cos$(\Theta_{\pi}^{\star}$).
}
\label{fig:models}       
\end{figure}
Recent analyses of the helicity amplitudes for resonance excitations on the neutron are 
given in \cite{Shresta_12,BnGa_n,SAID_n}. The comparison of the results from the 
different analyses (see e.g. \cite{BnGa_n}) reveals large discrepancies. Their absolute 
magnitudes differ up to two orders of magnitude, and for several states not even the sign 
is fixed. These problems arise from the lack of data for the $\gamma n\rightarrow n\pi^0$ reaction. 
The situation is sketched in Fig.~\ref{fig:models}. The bottom insert of the figure shows the 
available data bases for differential cross sections for all isospin channels \cite{Krusche_11}. 
Data are abundant for the $p\pi^0$, $n\pi^+$, and $p\pi^-$ final states, however, for $n\pi^0$ only a 
few scattered points have been measured. These points come from experiments in the early 1970s 
\cite{Bacci_72,Hemmi_73}, which could not even eliminate the background from photoproduction of 
$\pi^0$ pairs. Only the beam asymmetry $\Sigma$ has been measured precisely for this reaction 
by the GRAAL experiment \cite{DiSalvo_09}. 

In principle, the measurement of three reactions out of $\gamma p\rightarrow p\pi^0$, 
$\gamma p \rightarrow n\pi^+$, $\gamma n\rightarrow p\pi^-$, and $\gamma n\rightarrow n\pi^0$ 
should be sufficient according to Eq.~\ref{eq:iso} for an isospin decomposition \cite{Krusche_03}.
However, the predictions for the total cross section of $\gamma n\rightarrow n\pi^0$ by different
analyses agree only in the range of the $\Delta$(1232) resonance, for which the excitation of 
protons and neutrons is identical. In the second and third resonance region large discrepancies 
exist, although, as shown in the upper insert of Fig.~\ref{fig:models}, all models agree for
$\gamma p\rightarrow p\pi^0$ (because they have been fitted to the precise data available
for this reaction). So far, a unique partial-wave analysis is not possible for any of the different 
isospin channels (not enough observables have been measured yet). The analyses are thus model 
dependent and so is the prediction for the $n\pi^0$ channel. It is based on the data for
$\gamma n\rightarrow p\pi^-$, which has been measured by several experiments and recently 
with high accuracy by CLAS at Jlab \cite{SAID_n}. However, this reaction has different 
characteristics from $\gamma n\rightarrow n\pi^0$. The $p\pi^-$ final state has large contributions 
from non-resonant background terms since the $p\pi^-$ pair has electric charges and a large 
electric dipole moment to which the incident photon can couple. Non-resonant $t$-channel 
contributions (pion-pole, vector meson exchange etc.) are therefore much more probable than for 
$\gamma n\rightarrow n\pi^0$. Therefore the latter reaction is better suited for the extraction
of resonance properties from the $s$-channel contributions. The large differences between the
predictions from different models for the $\gamma n\rightarrow n\pi^0$ reaction mean of course that 
a measurement of this reaction could much better constrain the analyses. The lack of data 
for this channel is 
rooted in the specific experimental problems related to the measurement of an all-neutral 
(only photons and neutron) final state from quasi-free neutrons bound in the deuteron 
\cite{Krusche_11}. The present results close this gap in the data base. 

In this Letter, we report results from the first precise measurement of the total and differential
cross sections for the $\gamma d\rightarrow p(n)\pi^0$ and 
$\gamma d\rightarrow n(p)\pi^0$ reactions (in brackets: undetected spectator nucleon) 
in quasi-free kinematics. The experiment was performed at the Mainz MAMI accelerator
\cite{Kaiser_08}, which delivered the primary electron beam of 1.508~GeV (1.557~GeV)
energy (two different beam time periods were analyzed). 
The tagged photon beam was produced by bremsstrahlung of the electrons in a copper radiator 
of 10~$\mu$m thickness. The scattered electrons were momentum analyzed and
detected in the focal plane of the Glasgow-Mainz Tagged Photon Spectrometer \cite{McGeorge_08}
in order to tag the photon energies. The resulting photon beam   
had energies between $\approx$0.45~GeV and $\approx$1.4~GeV with an energy resolution
of $\approx$4~MeV. This beam impinged on liquid deuterium in cylindrical Kapton cells 
of $\approx$4~cm diameter and lengths of 4.72~cm (3.02~cm), mounted in the center 
of the detector such that the photon beam passed along their symmetry axis. The detector 
combined the Crystal Ball (CB) \cite{Starostin_01} and TAPS \cite{Gabler_94} electromagnetic 
calorimeters supplemented by charged-particle identification devices. 
The 672 triangular NaI(Tl) crystals of the CB covered the full azimuthal 
range for polar angles from 20$^{\circ}$ to 160$^{\circ}$, while the 366 hexagonally 
shaped BaF$_{2}$ crystals of TAPS were arranged as a forward wall placed 1.457~m 
downstream of the target at polar angles between $\approx$5$^{\circ}$ and
$\approx$21$^{\circ}$. The setup covered more than $95\%$ of the full solid angle. 
For the identification of charged particles, a Particle Identification 
Detector (PID) consisting of 24 plastic scintillator strips \cite{Watts_04} was mounted 
around the target and all crystals of TAPS had individual plastic scintillators in front. 
More details are given in  \cite{Oberle_13,Oberle_13a}, where the same data were analyzed 
for the production of pion pairs.   

Single $\pi^0$ production was analyzed in coincidence with recoil protons ($\sigma_p$) 
and in coincidence with recoil neutrons ($\sigma_n$). The identification of the photons,
protons, and neutrons was based on a combination of energy deposit in CB and in TAPS;
$\Delta E-E$ analysis from PID and CB; time-of-flight (ToF) versus energy from TAPS
and pulse-shape analysis from TAPS. The $\pi^0$ was identified from the invariant mass
of photon pairs. The reaction identification (i.e. elimination of events from double pion 
production, $\eta\rightarrow 3\pi^0$ decays etc.) was achieved with a missing-mass analysis 
and a condition on coplanarity of meson and recoil nucleon. Details about these analysis 
procedures, which resulted in very clean data samples, are given e.g. in 
\cite{Oberle_13,Oberle_13a,Zehr_12}; the details of the present analysis will be published 
elsewhere. The absolute normalization of the cross sections was calculated with the surface 
densities of the targets (0.231$\pm$0.005 nuclei/barn and 0.147$\pm$0.003 nuclei/barn, 
respectively), the incident photon flux, and the acceptance and detection efficiency 
of the detector. The photon flux was derived from the number of scattered electrons and 
the tagging efficiency, i.e., the number of correlated photons that passed the collimator, 
as described in \cite{Oberle_13,Oberle_13a,Zehr_12}. The detection efficiency was modeled 
with Monte Carlo simulations using the Geant4 code \cite{Geant4}. The detection efficiency 
for the recoil nucleon was additionally determined with an analysis of the 
$\gamma p\rightarrow p\eta$ and $\gamma p\rightarrow n\pi^0\pi^+$ reactions measured 
with a liquid hydrogen target. These results improved the precision in particular
for critical geometries such as the transition region between the CB and TAPS.
Effects from nuclear Fermi motion were removed with a kinematic reconstruction of 
the nucleon-meson final-state invariant mass $W$ as discussed in \cite{Krusche_11}.

As a final check, the inclusive cross section $\sigma_{\rm incl}$ was extracted. 
This analysis had no conditions for recoil nucleons, which may have been detected or not. 
If they were detected, they were ignored. Since in the energy range of interest, coherent 
production of $\pi^0$ mesons off the deuteron is only a negligible fraction of the total 
cross section \cite{Krusche_99}, the relation 
$\sigma_{\rm incl}(E_{\gamma})\approx\sigma_p(E_{\gamma}) +\sigma_n(E_{\gamma})$ 
must hold (reconstruction of $W$ is not possible without recoil nucleons, so that 
the Fermi-smeared versions as a function of incident photon energy had to be used). 
The agreement was excellent, putting stringent limits on systematic uncertainties
in the recoil nucleon detection.    

\begin{figure}[thb]
\centerline{\resizebox{0.50\textwidth}{!}{%
  \includegraphics{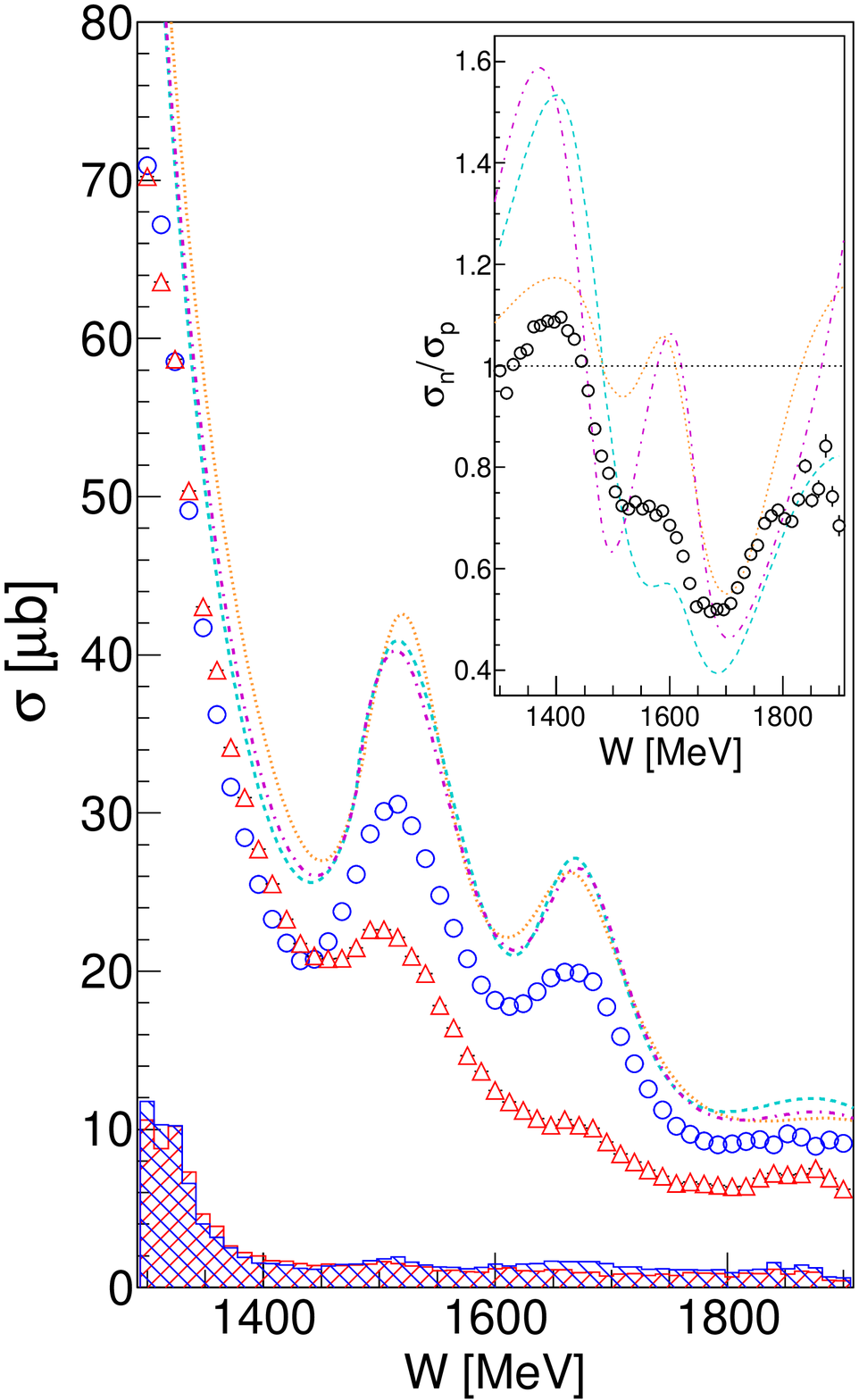}
  \includegraphics{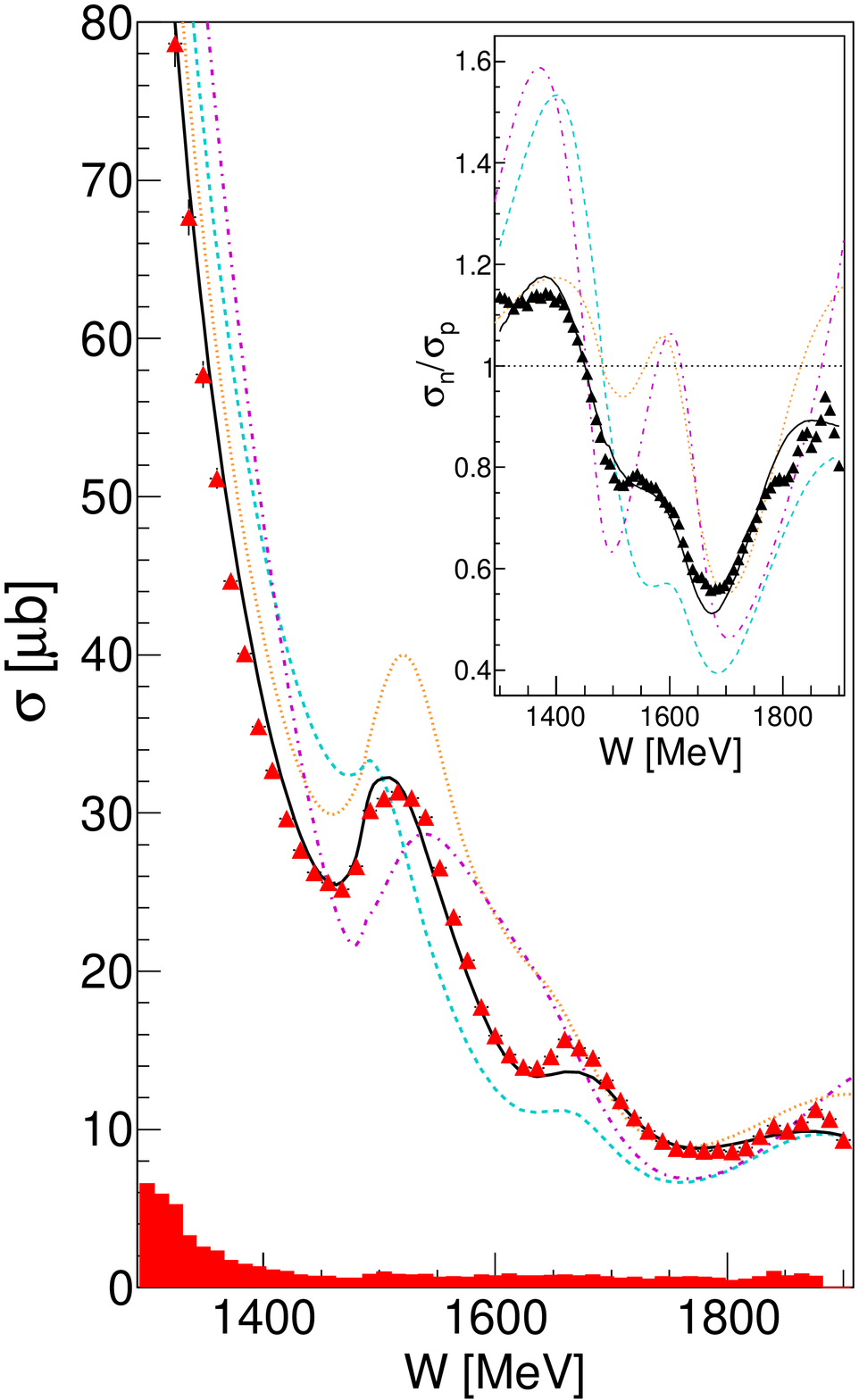}  
}}
\caption{Left-hand side: Total cross sections as functions of $W$ for $\pi^{0}$ production 
on quasi-free protons (open, blue circles) and on quasi-free neutrons (open, red triangles).
Curves: predictions for $\gamma p\rightarrow p\pi^0$ from the SAID multipole analysis 
\cite{SAID_new} dashed (cyan) line, the MAID unitary isobar model \cite{MAID_new} dotted
(orange), the BnGa analysis \cite{BnGa,BnGa_n} dash-dotted (magenta). Histograms at 
bottom represent systematic uncertainties (blue: proton, red: neutron). The insert compares 
the neutron/proton cross section ratio to the model predictions (same notation for curves).
Right-hand side: Total cross sections for $\gamma n\rightarrow n\pi^0$ (filled red triangles), 
i.e. quasi-free neutron data with FSI correction. Curves: predictions from same models as 
on left hand side, additionally (black solid) re-fit of BnGa model. Histogram at bottom: 
systematic uncertainty. Insert: ratio of corrected neutron cross section and SAID proton 
cross section.}
\label{fig:total}       
\end{figure}

\begin{figure}[thb]
\centerline{\resizebox{0.24\textwidth}{!}{%
  \includegraphics{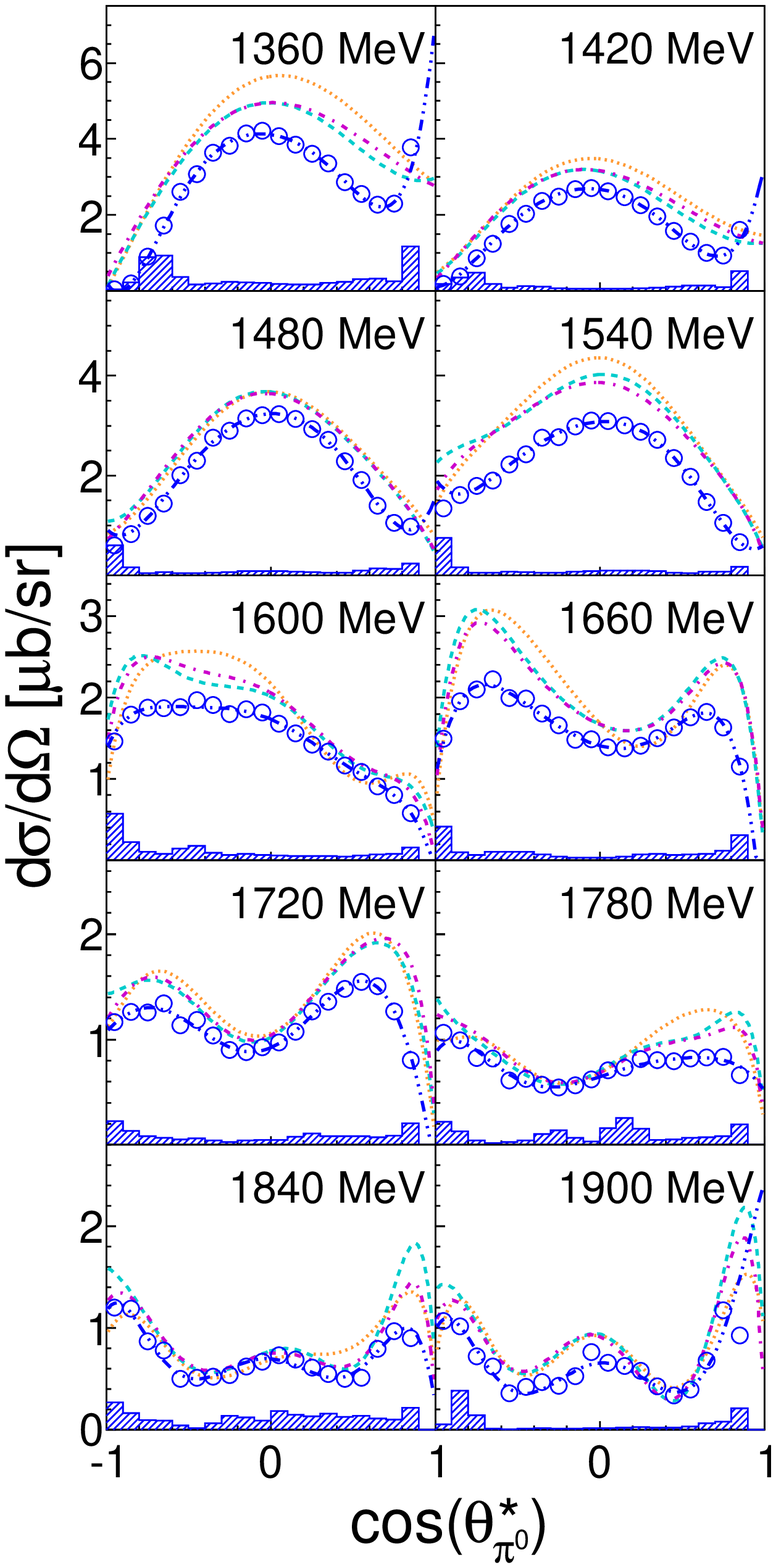}
}\resizebox{0.24\textwidth}{!}{%
  \includegraphics{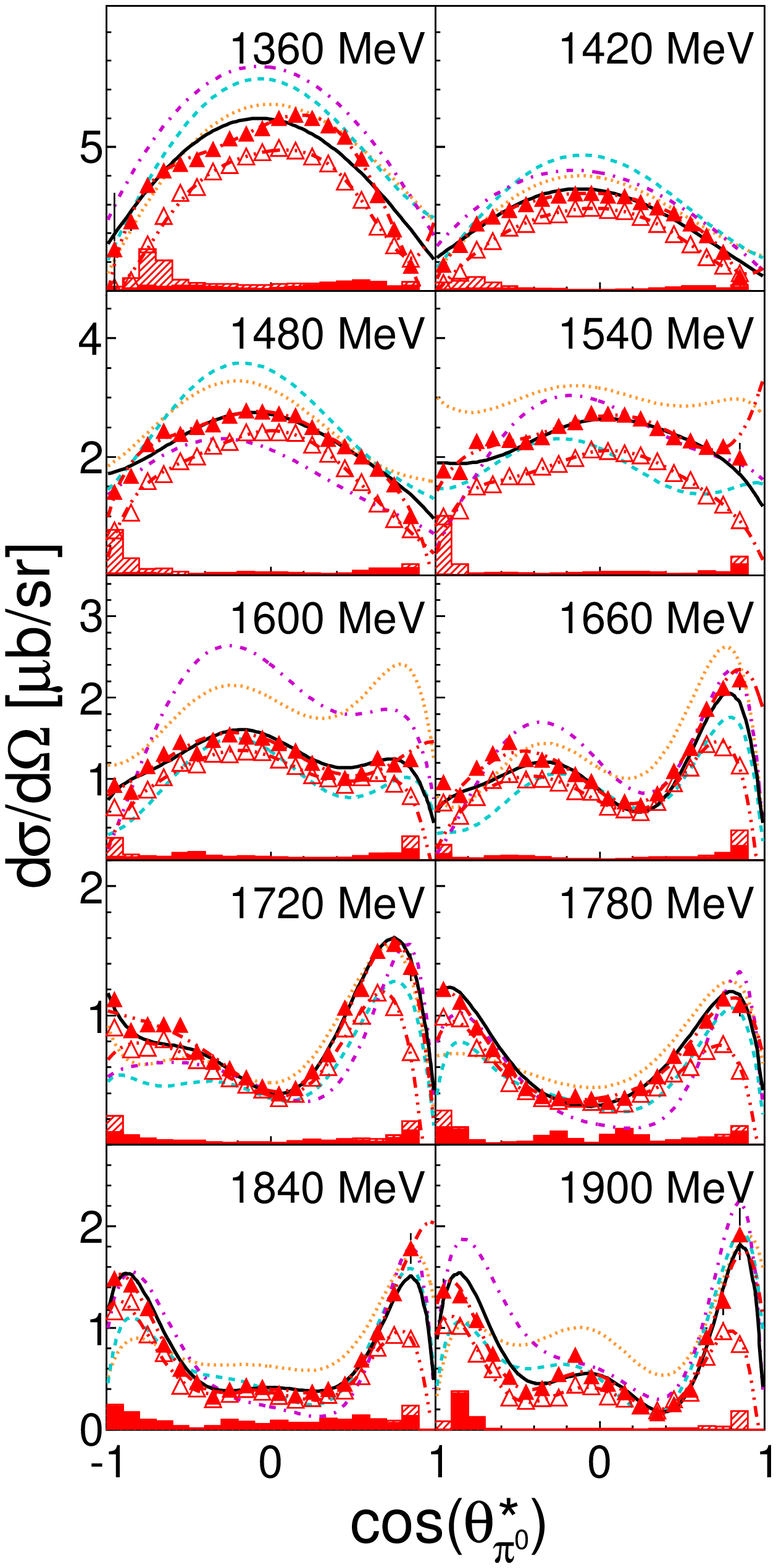}
}}
\caption{Angular distributions for $\pi^{0}$-photoproduction on the quasi-free 
proton (left two rows) and the quasi-free neutron (right two rows, filled symbols: 
FSI corrected, open symbols, no correction). Dash-dotted blue and red curves: 
fits to data with Eq.~\ref{eq:legendre}. 
Same coding for model curves as in Fig. \ref{fig:total}.}
\label{fig:diff}       
\end{figure}

The measured angular distributions (typical examples are summarized in Fig.~\ref{fig:diff})
were fitted with Legendre polynomials
\begin{equation}
\frac{d\sigma}{d\Omega} = 
\sum_{i=0}^{6} A_iP_i(cos(\Theta^{\star}_{\pi^0})) \ ,
\label{eq:legendre}
\end{equation}
and total cross sections were extracted from $\sigma =4\pi A_0$. They are compared 
in Fig.~\ref{fig:total} to predictions from the SAID partial-wave analysis \cite{SAID_new}, 
the MAID unitary isobar model \cite{MAID_new}, and the Bonn-Gatchina (BnGa) coupled-channel
partial-wave analysis \cite{BnGa,BnGa_n}.  
The left-hand side of the figure shows the measured proton and neutron cross sections
and the model results for the reaction off the free proton. The insert shows the neutron to proton 
cross-section ratios for data and models. A comparison of the quasi-free proton and neutron 
cross sections reveals that the peaks in the second- and third-resonance region of the nucleon 
are much suppressed for neutrons. For the third-resonance region this was predicted 
because the $F_{15}(1680)$ resonance has a much larger photon coupling for the proton than for 
the neutron, but for the second-resonance region predictions for the neutron/proton cross-section 
ratio vary widely.

The predictions for the free proton cross section are in close agreement because all models were 
fitted to the same data base. However, agreement with the quasi-free data is poor. This indicates 
that the quasi-free data are modified by nuclear effects such as final-state interactions (FSI) 
which are not included in the models. Similar effects had been previously observed and modeled 
for the quasi-free $\gamma d\rightarrow pp\pi^-$ reaction \cite{Tarasov_11}. A substantial 
suppression of the second-resonance peak for the inclusive $\gamma d\rightarrow \pi^0 np$ reaction 
with respect to the Fermi-smeared sum of the model predictions for $\gamma p\rightarrow p\pi^0$ 
and $\gamma n\rightarrow n\pi^0$ was already reported in \cite{Krusche_99}, but the inclusive data 
could not distinguish between FSI effects and a much smaller than predicted neutron cross section. 
The present data demonstrate that both effects contribute. A modeling of the FSI effects for 
$\pi^0$ production off the deuteron is not yet available, however, a comparison of the present 
quasi-free proton data to free proton data will allow a detailed test of future model results. 

An approximate correction for the nuclear effects was applied to the neutron data. It assumes 
that the effects are similar for quasi-free photoproduction of $\pi^0$ mesons off protons and 
off neutrons. For each angular distribution the ratio of the quasi-free proton data and the 
SAID results (which represent the average of all free proton data) was computed and applied 
to the neutron data as correction factors.
The total cross section for $\gamma n\rightarrow n\pi^0$ off free neutrons obtained this 
way is shown at the right-hand side of Fig.~\ref{fig:total} and compared to the model predictions. 
As already discussed, the latter deviate greatly and none of them is very close to the data 
(it seems the SAID predictions are closer than the other analyses). 

\begin{figure}[thb!]
\centerline{\resizebox{0.5\textwidth}{!}{%
  \includegraphics{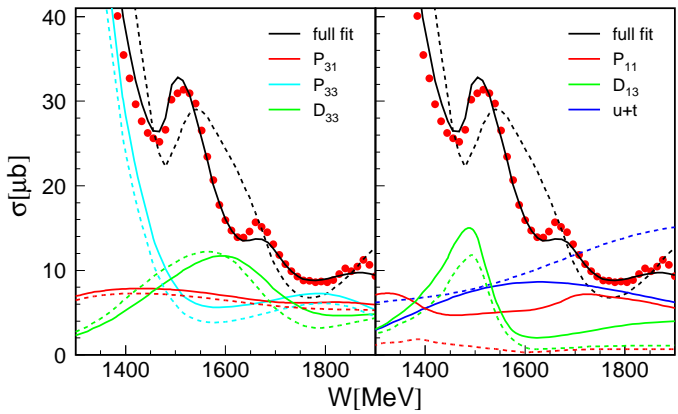}
}}
\caption{Results of BnGa fit. (Red) points: total cross section data, dashed curves: previous BnGa
results \cite{BnGa_n}, solid curves: re-fit including present data. Left hand side: partial waves
for $\Delta$-resonances (red: $P_{31}$, light blue: $P_{33}$, green: $D_{33}$). Right hand side:
$N^{\star}$ (red: $P_{11}$, green: $D_{13}$) and non-resonant background from $u-$ and $t$-channel
(blue).
}
\label{fig:pwa}       
\end{figure}

A re-fit of the
BnGa analysis (black solid curves in the figures), which included the present data, gets much 
closer to our results (and still describes also the previous data for the other reactions). 
For this re-fit the strength of background contributions (in particular vector-meson exchange) 
and several resonance couplings had to be modified. The impact of the data on this analysis is
demonstrated for some partial waves in Fig.~\ref{fig:pwa}. At the left hand side of the figure 
the total cross section and the contributions from strong, low-order resonant partial waves for
isospin $I=3/2$ $\Delta$-states ($P_{31}$, $P_{33}$, $D_{33}$) are shown. As expected, these partial
waves are not much affected because they are already sufficiently constrained by the results from the 
$\gamma p\rightarrow p\pi^0$ reaction. The situation is completely different for the isospin $I=1/2$
$N^{\star}$ states and the non-resonant backgrounds from $t$- and $u$-channel exchange shown at 
the right hand side of the figure. These amplitudes are not fixed by $\gamma p\rightarrow p\pi^0$
and their relative contributions to $\gamma n\rightarrow p\pi^-$ are much different. The resonant
$P_{11}$ partial wave (e.g. with contributions from the much discussed $P_{11}$(1440) 
`Roper'-resonance and the $P_{11}$(1710) state) changes drastically over the full energy range.
Above 1.6 GeV also the contributions from D$_{13}$-states (the photon coupling of the $D_{13}$(1700)
changes sign) and from non-resonant backgrounds are strongly influenced. Detailed results from 
the BnGa re-fit will be published elsewhere. 

In summary, for the first time angular distributions for the quasi-free reaction 
$\gamma n\rightarrow n\pi^0$ using a deuterium target were measured. 
The simultaneously measured quasi-free reaction off the proton establishes a data base 
for the investigation of nuclear effects such as FSI processes. The comparison of 
neutron and proton cross sections indicates the differences in resonance excitations 
off protons and off neutrons. A comparison of different model predictions for  
$\gamma n\rightarrow n\pi^0$ demonstrates that input data from the three other isospin
channels alone ($p\pi^0$, $p\pi^-$, $n\pi^+$ final states) cannot sufficiently constrain the 
isospin structure of pion photoproduction in the analyses. The experimental results for the
$\gamma n\rightarrow n\pi^0$ reaction are essential for the determination of the neutron 
helicity couplings of $N^{\star}$ resonances. A re-fit of the BnGa model shows the large 
impact of the new results on partial waves related to $N^{\star}$ excitations. 
The next step towards reliable neutron couplings requires re-analyses 
also in the framework of the other models in order to test whether the results
from the different approaches converge and, on the experimental side, also measurements
of polarization observables for $\gamma n\rightarrow n\pi^0$.

We wish to acknowledge the outstanding support of the accelerator group 
and operators of MAMI. We thank A.~Sarantsev and V. Nikonov for providing 
the results of the updated Bonn-Gatchina fit prior to publication.
This work was supported by Schweizerischer Nationalfonds
(200020-132799,121781,117601,113511), Deutsche
Forschungsgemeinschaft (SFB 443, SFB/TR 16), DFG-RFBR (Grant No. 05-02-04014),
UK Science and Technology Facilities Council, (STFC 57071/1, 50727/1), 
European Community-Research Infrastructure Activity (FP6), the US DOE, US NSF, and
NSERC (Canada).

\end{document}